# Investigation of the topography-dependent current in conductive AFM and the calibration method

*Chunlin Hao,[1] Hao Xu,[1] Shiquan Lin,[2] Jinmiao He,[1] Bei Liu,[1] Yongqiu Li,[1] Jiantao Wang,[1] Yaju Zhang[1],\*, and Haiwu Zheng[1],\**

[1]Henan Province Engineering Research Center of Smart Micro-nano Sensing Technology and Application, School of Physics and Electronics, Henan University, Kaifeng 475004, P. R. China

[2]Beijing Institute of Nanoenergy and Nanosystems Chinese Academy of Sciences, Beijing 100083, P. R. China

E-mail:

\*Yaju Zhang (40070002@vip.henu.edu.cn)

\*Haiwu Zheng (zhenghaiw@ustc.edu)

# ABSTRACT

The topography and the electrical properties are two crucial characteristics in determining roles and functionalities of materials. Conductive atomic force microscopy (CAFM) is widely recognized for its ability to independently measure the topology and conductivity. The increasing trend towards miniaturization in electrical devices and sensors has encouraged an urgent demand for enhancing the accuracy of CAFM characterization. However, the possibility of topography interference with the measured current during CAFM scanning leads to an inaccurate estimation of the sample's conductivity. Herein, we investigated the topography-dependent current originating from variation in capacitance between the probe and sample during CAFM testing. Based on the linear dependence between the current and the first derivative of height derived from topographic mapping, the calibration method has been proposed to eliminate the current error that is attributed to the variation in height on sample surfaces. This method is evaluated on one-dimensional ZnO nanowire, two-dimensional (2D) $NbOI_2$ flake, and biological lotus leaf, further demonstrating the feasibility and university of this method. This work effectively addresses the challenge of topographic crosstalk in CAFM characterization, which provides significant benefits for research on demanding high-accuracy CAFM measurements.

With advances in nanoscience and nanotechnology, atomic force microscopy (AFM) has been further developed to include new applications for characterizing material's electrical properties, such as Kelvin probe force microscopy (KPFM),[1] piezoresponse force microscopy (PFM),[2] and CAFM.[3] Among these techniques, CAFM provided a substantial advantage over the Nobel Prize-winning Scanning Tunneling Microscope by allowing the independent observation of topography and current signals.[4] Specifically, the measurement of cantilever's deflection is used to acquire the topographic map, while the current map is generated through the recording of current using a preamplifier.[5] This decoupling of topography and current signal is conductive to simultaneous mapping of surface topography and measurement of electrical properties with high spatial resolution.

The universal existence of topographic fluctuation in artificially constructed and biological materials, such as nanowires[6,7], 2D materials[8,9], lotus leaf[10,11], may interfere with the electrical properties, which results in the deviation of the actual conductivity. Previous study has demonstrated that topography fluctuations have an influence on the measured current during CAFM testing to some degree, arising from the variation of contact area caused by the force between the probe and the sample.[12] This situation makes it difficult to determine whether any variation in current from one location to another is due to the difference in topography or conductivity, which may cause the misjudgment of property and thus delayed use of materials. On the other hand, based on the potential difference between the sample and the probe inducing capacitance,[13] the current is generated by the variation in capacitance during the vertical movement of the probe.[14] These phenomena of topography- and capacitance-based current interfere with the determination of electrical conductivity from current measurements. Therefore, it is imperative to conduct a comprehensive study for decoupling the sample's topography and measured current during CAFM testing.

In this study, topography-dependent current has been investigated in CAFM testing on the mica substrate with uneven surface, which is generated by the variation in capacitance from the fluctuation in probe height. The interference between the measured current and the topography leads to confusing conductivity. To decouple the measured current resulting from topographic fluctuation, the definitive relationship between the current and the sample topography is determined experimentally and theoretically. Based on this correlation, a comprehensive method is proposed to eliminate the topography-based current, which is further evaluated on a-$Ga_2O_3$/ZnO nanowire array, 2D $NbOI_2$ flake and lotus leaf. This work will offer substantial advantages for high-precision CAFM testing in nano research fields, which will open up new possibilities for eliminating topographic crosstalk in scanning probe microscopy and greatly enrich the research connotation of microspectroscopy based on CAFM.

## RESULTS AND DISCUSSION

Figure 1a illustrates the sample preparation process for a square pit on the mica substrate through photolithography and lift-off techniques. The detailed preparation process is described in the Experimental section. During the CAFM test, the silver paint acted as positive electrode with a bias of 9 V, while keeping the probe grounded. "Trace" and "Retrace" are two distinct channels that can be obtained from opposite scanning directions. Each line is scanned from left to right (Figure 1b-d), known as trace; and then from right to left (Figure 1e-g), known as retrace.

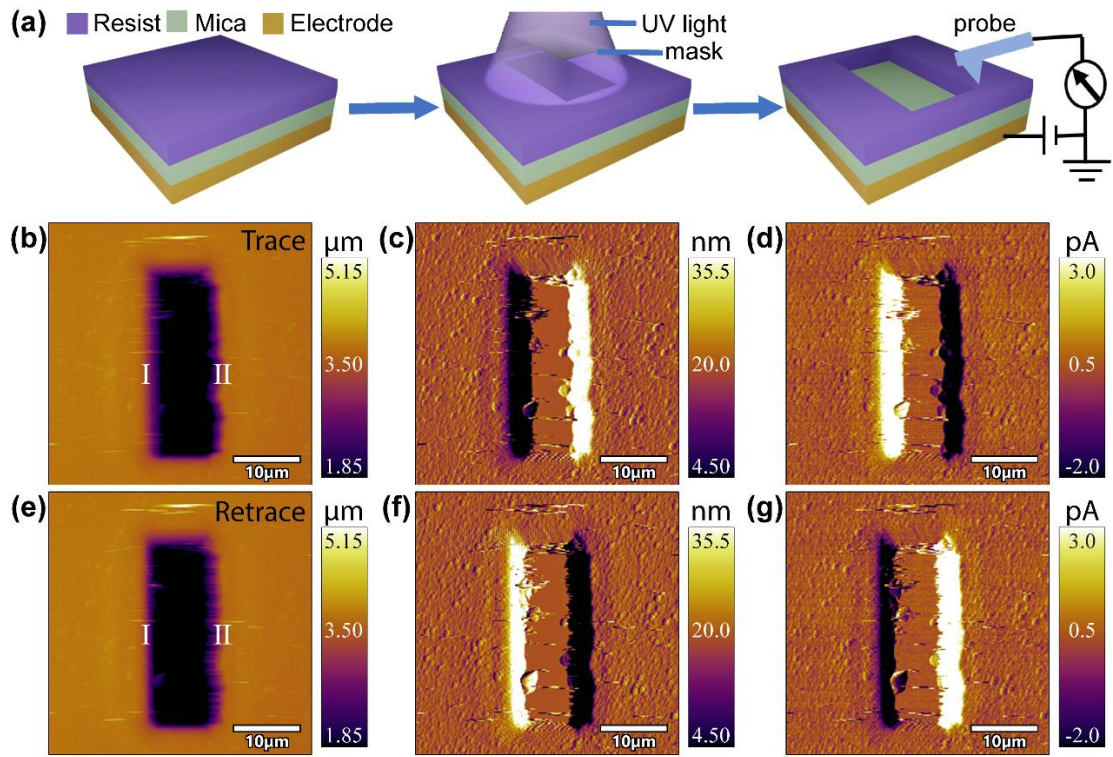

**Figure 1.** (a) Schematic diagram of the sample preparation process and CAFM testing. (b-d) Trace of topography, deflection, and current. (e-g) Retrace of topography, deflection, and current.

As can be seen from Figure 1b,e, regions (I) and (II) are two edges of the pit with significant height difference. These two identical images indicate that the scanning direction does not influence topography results. It is noteworthy that the trace and retrace of both deflection (Figure 1c,f) and current channels (Figure 1d,g) demonstrate meaningfully discriminative. The trace of deflection, representing both the force exerted by the probe on the sample and the height gradient, decreases at region (I) whereas increases at region (II). The retrace of deflection shows opposite pattern at locations (I) and (II). Correspondingly, the opposite current bursts appear during the trace and retrace at locations (I) and (II), meaning that generation of current is associated with variation in force and height between the probe and the mica. One of the typical examples is to induce current bursts in the step edges of topological insulator $Bi_2Te_3$, which is attributed to the variation in the force applied by the probe.[12] On the contrary, there is negligible current in insulating mica, regardless of any variation in the force exerted by the probe. It is speculated that the bursts of current should be ascribe to height gradient.

To explore the origin of current variation, we exerted bias voltages on the capacitor consisting of the CAFM probe and mica with a smooth surface (Figure 2a). Four different bias voltages (9, 4.5, -4.5, and -9 V) are applied in regions (I), (II), (III), and (IV), respectively, as illustrated in Figure 2b. By adjusting the height of the probe (referring to Figure S1 and the Experimental section), trace scanning (Figure 2c) is utilized to acquire topography information of the mica, and during retrace scanning (Figure 2d), the probe scans at various predetermined heights. The roughness of mica on the order of 166.142 pm is much less than the probe's height fluctuation up to 5 μm, thus the influence of the roughness on the measured current can be considered negligible. As can be seen from Figure 2e, when the height trace remains constant, the current keeps steady under different predetermined bias voltages. By contrast, once the height retrace varies, the current bursts occur (Figure 2f). The current exhibits a direction inversion during the ascent and descent processes. Additionally, the current retrace displays clearly distinguishable in regions (I), (II), (III), and (IV), corresponding to four applied bias voltages.

Figure 2g depicts cross-sectional lines extracted from Figures 2c-f. During the retrace scanning, there is no contact between the probe and mica in section AD. As the probe's height increases in section AB, the negative current generates in region (I) under the bias of 9 V. When the probe reaches the highest point and remains constant in section BC, no current exists. As the probe descends in section CD, the positive current produces. Likewise, bursts of currents are present in regions (II), (III), and (IV) where the height of probe varies. The higher bias voltage, the higher magnitude of current. The reversing in the direction of current arises in regions (III) and (IV) under negative bias voltage. These results reveal that the generation of current closely relates to variation in height of probe other than conduction current[12] and triboelectricity[15,16] induced by the contact force between the probe and sample.

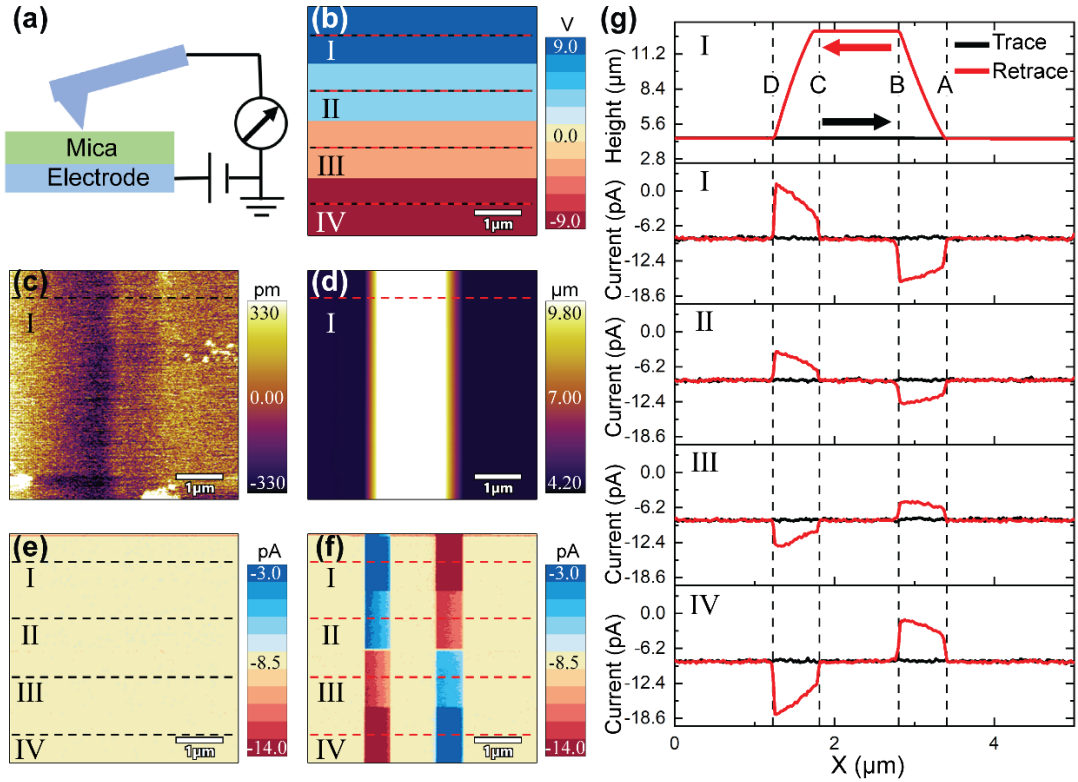

**Figure 2.** (a) Schematic diagram of CAFM testing. (b) The distribution of applied bias voltage during the scanning test. (c,d) Trace and retrace of height channel after program modification. (e,f) Trace and retrace of the current channel after program modification. (g) Cross-sectional lines extracted from Figure 2c-f. Black lines represent trace scanning and red lines represent retrace scanning.

Figure 3 depicts a schematic diagram of the mechanism of current generation. When the distance between the probe and conductive sample is constant, the total current measured by CAFM is composed of two components, as stated in Equation 1:

$$I_{total} = \frac{V}{R} + I_{offset} = I_{cond} + I_{offset} \tag{1}$$

where $I_{offset}$ is the current offset of the instrument and remains constant throughout the test. $I_{cond}$ is related to the sample's conductivity (as shown in Figure 3a). If the sample is an insulator (Figure 1 and 2), $I_{cond}$ is negligible.

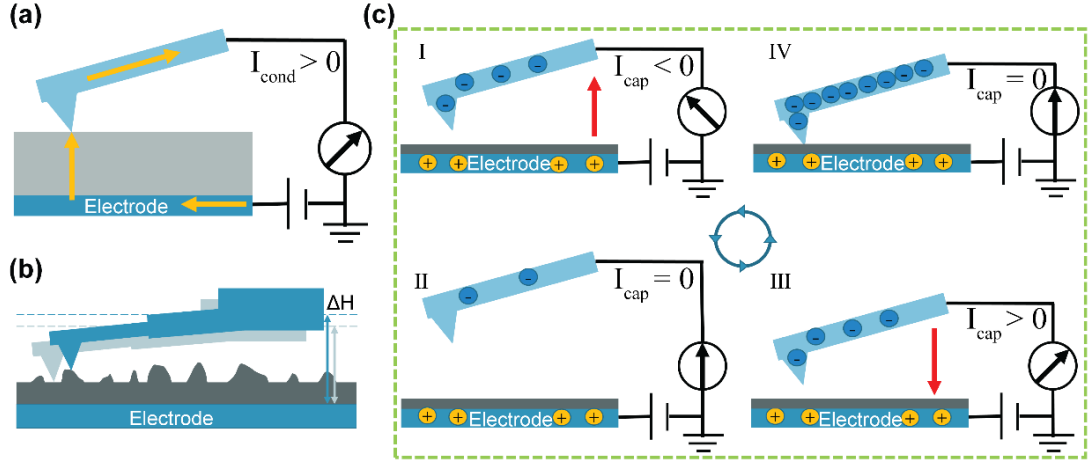

**Figure 3.** (a) Schematic diagram of positive conduction current passing through the CAFM and sample. (b) The variation in the probe's height during the scanning process. (c) The schematic diagram of current generation mechanism.

When the distance between the probe and insulating sample is varied, the total current measured by CAFM includes $I_{cap}$, which is generated by the variation in capacitance between the probe and the sample, as shown in Equation 2:

$$I_{total} = I_{cap} + I_{offset} \tag{2}$$

As shown in Figure 3b, the upper plate of the capacitor is composed of the probe, holder, and instrument.[17,18] As the probe scans over the uneven surface of the sample, the variation in distance between the probe and sample ($\Delta H$) leads to variation in capacitance (Figure 3b). The equivalent capacitance satisfies Equation3:

$$C = \frac{\varepsilon S}{H} = \frac{Q}{U} \tag{3}$$

where $C$ represents the equivalent capacitance, $S$ is the equivalent plate area, $H$ is the equivalent spacing between plates, $Q$ is the charge on each plate, $U$ is the bias voltage applied on the sample, and $\varepsilon$ denotes the dielectric constant. Variation in $H$ induces corresponding variation in $Q$, which generates a current, as shown in Figure 3c.

Specifically, as the probe's height increases, the capacitance decreases (Figure 3c(I)), which corresponds to section AB in Figure 2g. Negative charges flow out from the probe and a negative current generates by applying positive bias to the bottom electrode, corresponding to Figures 2g(I) and (II). Conversely, a positive current generates by applying negative bias to the bottom electrode, corresponding to Figures 2g(III) and (IV). As the probe reaches the highest point and remains constant, there is no variation in capacitance, leading to no current (Figure 3c(II) corresponding to section BC in Figure 2g).

As the probe descends, the capacitance increases (Figure 3c(III)), which corresponds to section CD in Figure 2g. Negative charges flow into the probe resulting in a positive current by applying positive bias to the bottom electrode, corresponding to Figures 2g(I) and (II). Conversely, the current becomes negative by applying negative bias to the bottom electrode, corresponding to Figures 2g(III) and (IV). As the probe maintains the constant height where it touches the sample (Figure 3c(IV)), there is no variation in height, thus no current generates.

According to the above-mentioned current generation mechanism, the quantitative relationship between current, variation in height and bias is further established. A series of force curves including the height, current and bias dependent on time were conducted at a single point. As can be seen from Figure 4a, the height of probe is increased by 5 μm after the probe touching the mica at point A, and then is adjusted according to triangular waves with different frequencies and amplitudes. The determined amplitudes and frequencies of the triangular waves were 2 μm and 6.25 Hz (blue region), 2 μm and 12.5 Hz (red region), 4 μm and 6.25 Hz (green region), respectively. Owing to the variation in height of the probe, square-wave current signals with the same frequency are generated. The higher slope of height, the higher amplitude of current.

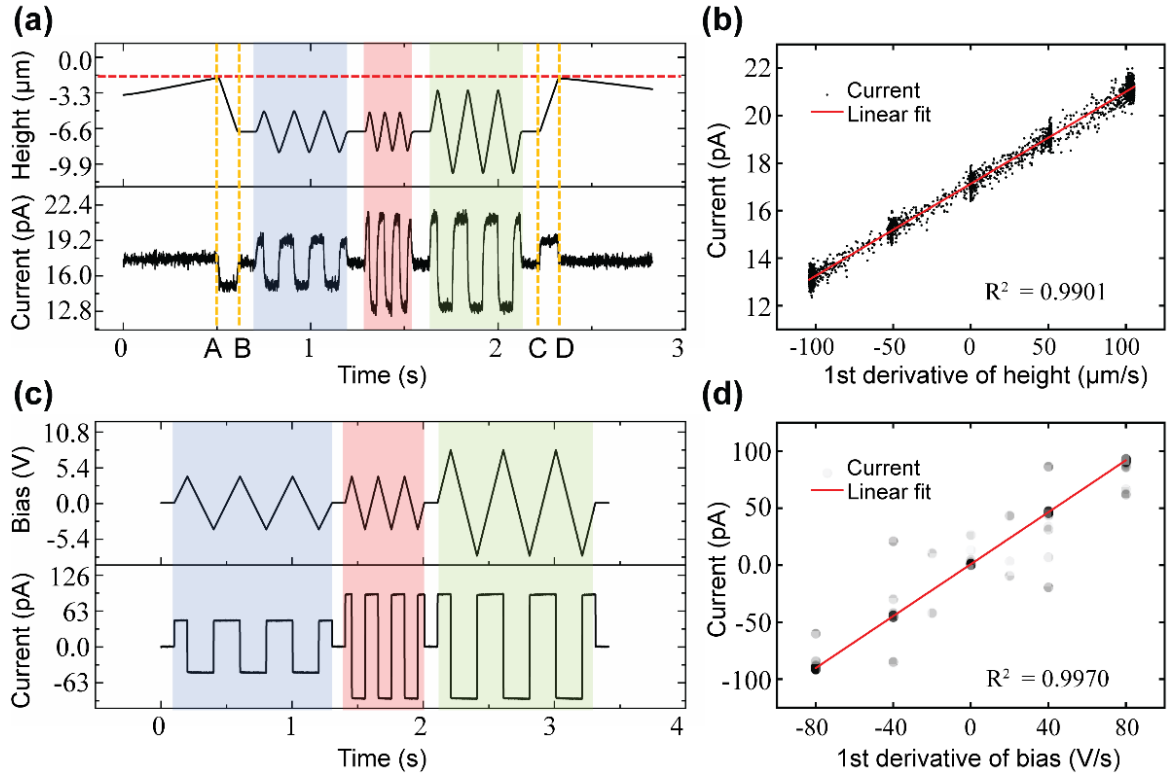

**Figure 4.** (a) The force curve depicting variation in probe height and current over time. (b) Correlation between the current and first derivative of height over time. (c) The force curve of variation in bias voltage and current over time. (d) Correlation between bias voltage and the first derivative of height over time.

Figure 4b demonstrates that there is a strong linear correlation (the coefficient of determination $R^2$ is 0.9901) between the current and the first derivative of height with respect to time d$H$/d$t$ extracted from the BC section, expressed as:

$$I_{cap} \propto \frac{dH}{dt} \tag{4}$$

The five distinct clusters are determined by the five pre-set slopes of height and the scattered dots among different clusters are derived from the continuous changes among the five pre-set slopes of height in the test. When the slope of height is zero, the measured current of 17 pA represents the current offset of the instrument.

The applied voltage is another parameter that influences the magnitude of current. As can be seen from Figure 4c, the determined amplitudes and frequencies of bias voltages are 4 V and 2.5 Hz (blue region), 4 V and 5 Hz (red region), 8 V and 2.5 Hz (green region), respectively. As the bias voltage varies, square-wave current signals with the same frequency are generated. The higher slope of bias voltage determines the higher amplitude of current.

Figure 4d represents the linear fit of the current and voltage extracted from Figure 4c, the current still exhibits a strong linear dependence on the first derivative of the bias voltage with respect to time $dU/dt$ ($R^2 = 0.997$), expressed as:

$$I_{cap} \propto \frac{dU}{dt} \tag{5}$$

Based on the aforementioned results, the relationships between the current and height as well as voltage are derived during the CAFM testing. The following will present theoretical derivation to validate Equation 4 and 5. We model a simple parallel-plate capacitor consisting of the probe and the sample. The current generated by the capacitor ($I_{cap}$) can be expressed as follows:

$$I_{cap} = \frac{dQ}{dt} = \frac{d\frac{\varepsilon S U}{H}}{dt} = \frac{\partial Q}{\partial U}\frac{dU}{dt} + \frac{\partial Q}{\partial H}\frac{dH}{dt} = \frac{\varepsilon S}{H}\frac{dU}{dt} - \frac{\varepsilon S U}{H^2}\frac{dH}{dt} \tag{6}$$

When the voltage remains constant and the height fluctuation is much less than $H$ ($\Delta H \approx 0$), $H^2$ can be considered approximately constant. The relationship between $I_{cap}$ and $dH/dt$ corresponds to Equation 4 and is expressed as follows:

$$I_{cap} = -\frac{\varepsilon S U}{H^2}\frac{dH}{dt} \tag{7}$$

The current is directly proportional to $U$ when $dH/dt$ remains constant and non-zero, which is consistent with Figure 2f.

When the height of the probe remains constant, the relationship between $I_{cap}$ and $dU/dt$ corresponds to Equation 5 and is expressed as follows:

$$I_{cap} = \frac{\varepsilon S}{H}\frac{dU}{dt} \tag{8}$$

The total measured current ($I_{total}$) can be expressed as:

$$I_{total} = I_{cap} + I_{cond} + I_{offset} \tag{9}$$

To accurately measure the conductivity of the sample ($I_{cond}$), it is imperative to explore the calibration method for eliminating $I_{cap}$. The calibration approach A is proposed by superimposing and averaging the positive and negative $I_{cap}$ generated during trace and retrace scanning according to Equation 7. Take region (I) from Figure 1 as an example to validate the feasibility of the calibration approach A. The direction of trace $I_{cap}$ when the probe descends at region (I) ($dH/dt < 0$) is opposite to the retrace $I_{cap}$ when the probe ascends at region (I) ($dH/dt > 0$).

Figure 5a illustrates the superimposing and averaging of the trace and retrace of current. Compared to Figure 1d and g, the $I_{cap}$ experiences some degree of reduction where the height between the probe and sample varies. The incomplete elimination of $I_{cap}$ originates from the slight discrepancy in height between the trace and retrace path (see Figure S2 for more information).

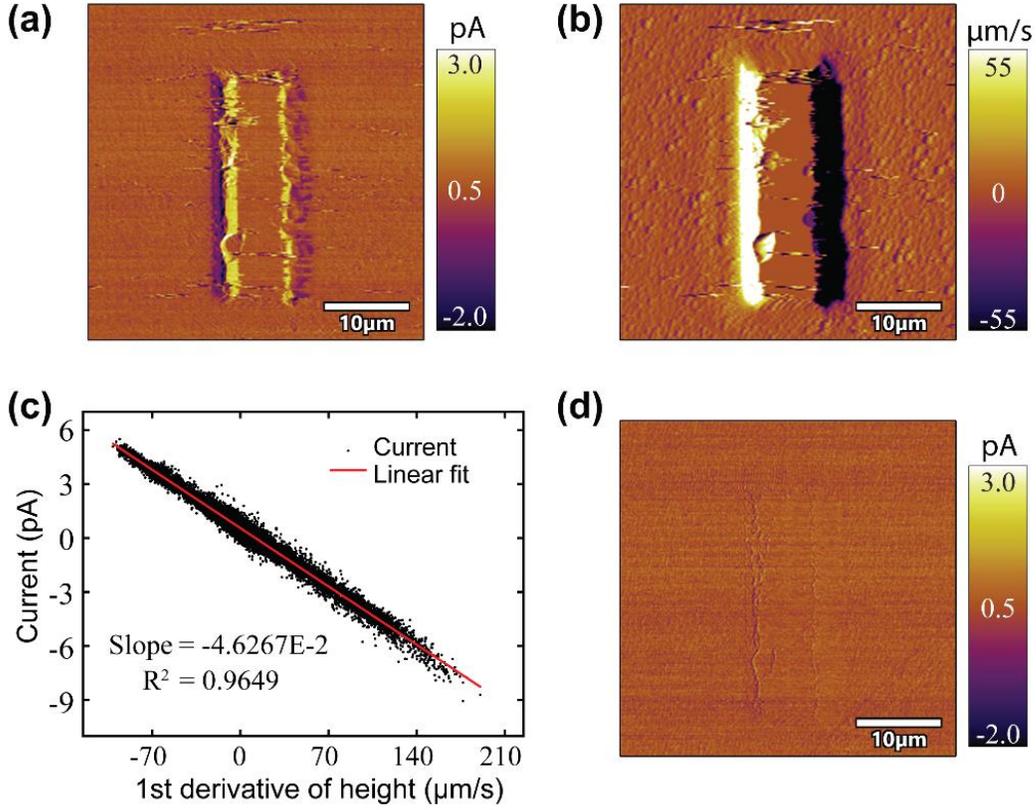

**Figure 5.** (a) Calibration result by superimposing and averaging data from Figure 1d,g. (b) The first derivative of data from Figure 1e with respect to time. (c) Linear fit of current and d$H$/d$t$ derived from Figure 1g and 5b, respectively. (d) Calibration result according to Equation 11.

An alternative calibration approach B is introduced based on Equation 7 and Equation 9, expressed as:

$$I_{cond} + I_{offset} = I_{total} - I_{cap} = I_{total} - \frac{dI_{cap}}{dH'}H' \tag{10}$$

where $H'$ represents the first derivative of height with respect to time (d$H$/d$t$).

Since the conductive current ($I_{cond}$) of mica is negligible, d$I_{cap}$/d$H'$ equals to d$I_{total}$/d$H'$, Equation 10 can be expressed as:

$$I_{offset} = I_{total} - \frac{dI_{total}}{dH'}H' \tag{11}$$

Figure 5b depicts the derivative of height extracted from Figure 1e with respect to time ($H'$), referring the Experimental section for detail. The $H'$ dependence of $I_{total}$ extracted from Figure 1g is depicted in Figure 5c. The coefficient of determination $R^2$ is 0.9649, indicating a strong linear correlation between $H'$ and $I_{total}$.

Figure 5d is derived by substituting $I_{total}$, $H'$, and $dI_{total}/dH'$ into Equation 11. The $I_{cap}$ is almost entirely eliminated, implying that the calibration method B is applicable for insulator materials ($I_{cond}=0$). For conductive samples, $I_{cond}$ is not equal to 0, it is essential to further verify the feasibility of calibration method B.

Taking the standard sample as an example, it is comprised of discontinuous $SiO_2$ as top layer and Si as sublayer (more information in the Experimental section), as shown in Figure 6a. As can be seen from the retrace of the sample's topography shown in Figure 6b, the height difference between the Si (dark region) and $SiO_2$ (bright region) is approximately 100 nm, which causes the generation of $I_{cap}$. When the probe makes contact with $SiO_2$ (Figure 6a(I)), no current is generated ($I_{cond}=0$) under 9 V bias. When the probe touches Si (Figure 6a(II)), charges flow through the Si and then are captured by the defects located at the resin/glass interface, inducing non-zero current ($I_{cond}>0$) under 9 V bias. In this case, the existence of the difference in height and conductivity between Si and $SiO_2$ induces the generation of $I_{cap}$ and $I_{cond}$, respectively.

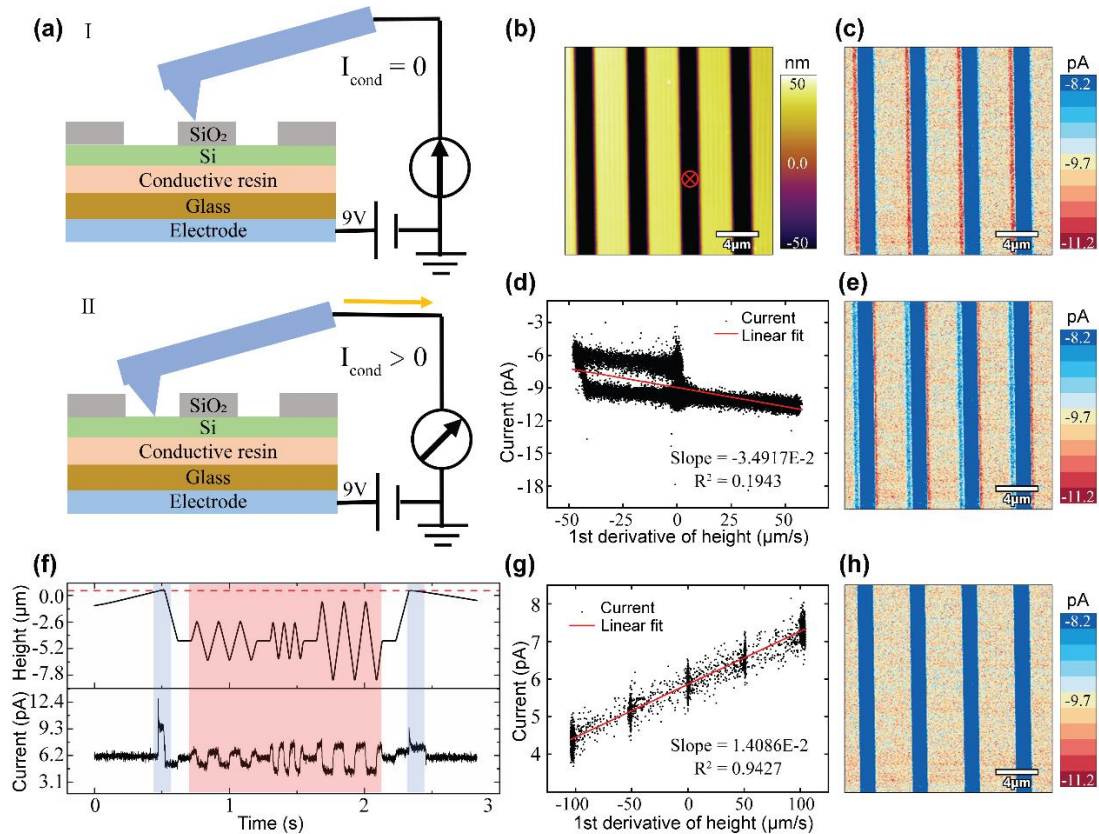

**Figure 6.** (a) Schematic diagrams of the probe making contact with Si and SiO$_2$ during the scanning, respectively. (b,c) Retrace of height and current. (d) Linear fit of $I_{total}$ and $H'$ derived from Figure 6b and 6c, respectively. (e) Calibration result according to Equation 11 by extracting data from Figure 6b-d. (f) The force curve derived at the red mark in Figure 6b. (g) Linear fit of the red section in Figure 6f. (h) Calibration result according to Equation 11 by extracting data from Figure 6b,c,g.

There is the presence of apparent current when the probe contacts with Si, as represented by the dark blue region in Figure 6c. As the height of probe increases and the probe contacts SiO$_2$, negative current ($I_{cap}$<0) generates, corresponding to the red region in Figure 6c. The occurrence of $I_{cap}$ leads to errors in measuring sample conductivity. The calibration method B represented by Equation 11 was applied to eliminate the $I_{cap}$. Figure 6d demonstrates the weak linear relationship between the $I_{total}$ and $H'$ with $R^2$ of only 0.1943. The $I_{cap}$ is calculated to be non-zero according to Equation 11 by extracting $H'$, $I_{total}$, and $dI_{total}/dH'$ from Figure 6b,c,d, respectively, corresponding to the light blue region in Figure 6e. The above-mentioned analyses confirm that the calibration method B is no longer applicable for non-zero $I_{cond}$ in response for the nonlinear relationship of $I_{cond}$ and $H'$ and the unequivalence between $dI_{total}/dH'$ and $dI_{cap}/dH'$.

Another calibration method C is introduced by updating $dI_{total}/dH'$ extracted from the force curve according to Equation 11. A force curve depicted in Figure 6f is derived at the point with red in Figure 6b. When there is no contact between the probe and Si ($I_{cond} = 0$), the $H'$ dependence of $I_{total}$ can be extracted from the red section, as shown in Figure 6g. The coefficient of determination $R^2$ with 0.9427 implies the strong linear correlation between $H'$ and $I_{total}$. Figure 6h is derived by substituting $H'$, $I_{total}$, and $dI_{total}/dH'$ (extracted from Figure 6b,c,g, respectively) into Equation 11. It demonstrates two kinds of conductivity distributions corresponding to the Si and $SiO_2$ regions, implying that $I_{cap}$ has been eliminated and the calibration method C is applicable for conductive materials.

The accuracy of updating $dI_{total}/dH'$ from the force curve other than topographic mapping has been examined in Figure S3. The margin of error less than 5% implies the calibration method C being reliable and universal for functional materials with different conductivity.

The feasibility of calibration method C is evaluated on artificially constructed and biological materials with different dimensions: a-$Ga_2O_3$/ZnO nanowire array[6,7,19], 2D $NbOI_2$ flake[8,9] and lotus leaf[10,11]. The topography and crystalline structure characterizations and corresponding schematic diagrams of CAFM tests are depicted in Figure S4.

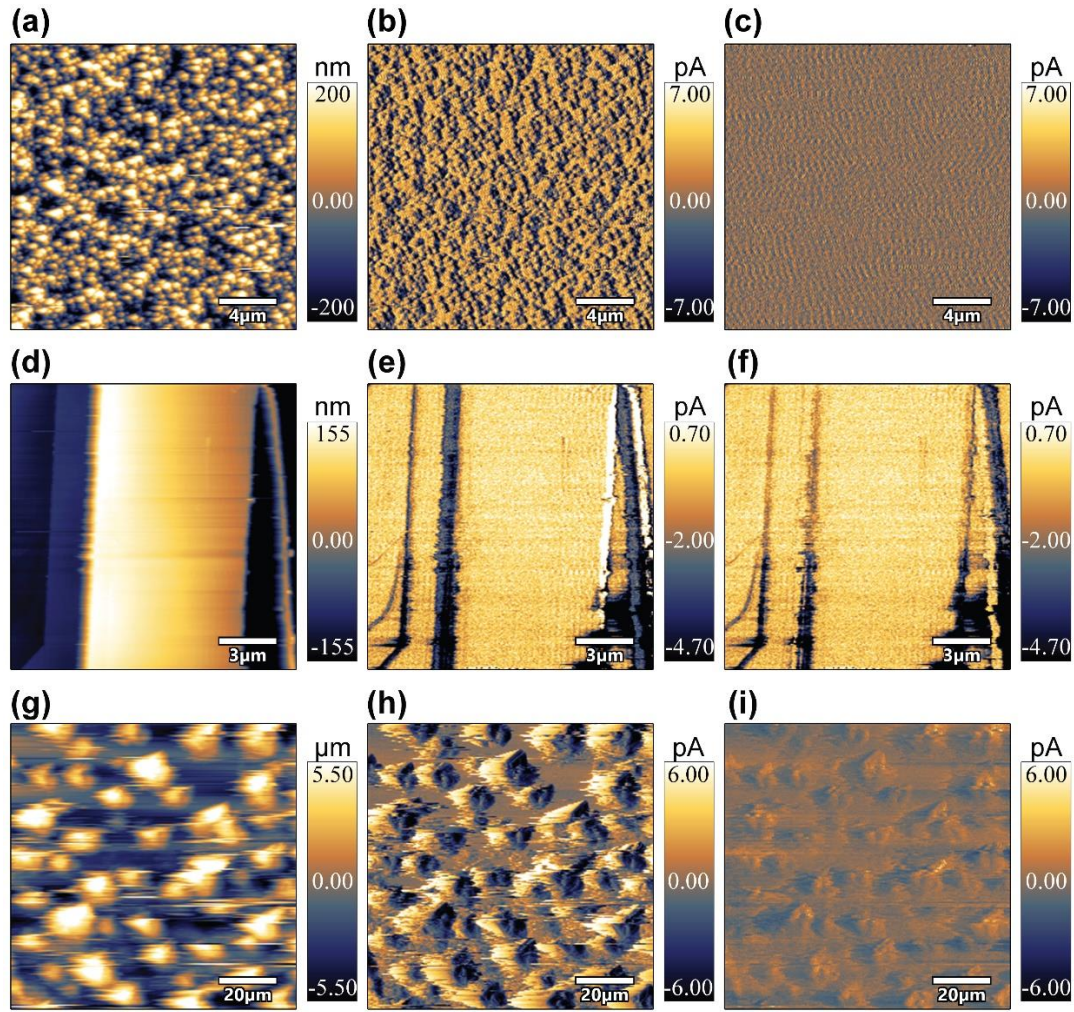

**Figure 7.** Topography retraces, current retraces and calibration results of (a-c) a-$Ga_2O_3$/ZnO nanowire array under 10 V bias, (d-f) 2D $NbOI_2$ flake under -10 V bias, and (g-i) lotus leaf under 10 V bias.

Figure 7a,b demonstrate the topography and the corresponding current distribution of the ZnO nanowire array. The topographic roughness of ZnO nanowire array leads to the presence of error, that is non-zero $I_{cond}$. By introducing calibration method C, the calibration current $I_{cond}$ is evenly distributed in Figure 7c and S5, implying no significant differences in actual conductivity. Likewise, the height difference at the step edges of 2D $NbOI_2$ flake causes the existence of positive and negative currents, as shown in Figure 7d,e. The calibration result in Figure 7f and S5 shows

that the positive current is effectively eliminated and residual current ($I_{cond} < 0$) results from the applied negative bias voltage. The calibration method C is further verified to exclude the influence of large height difference in superhydrophobic microstructure (Figure 7g) on measured current of lotus leaf (Figure 7h and S5) and then contributes to calibrating actual conductivity in Figure 7i and S5.

## CONCLUSIONS

In summary, we investigated the topography-dependent current for CAFM characterization on mica substrate with uneven topography. The current error $I_{cap}$ is generated due to the variable capacitance from the variation in heigh between the sample and the probe. Based on the strong linear correlation between the current and the first derivative of height or bias voltage, an extended calibration method is feasible for eliminating $I_{cap}$ by updating the slope ($dI_{total}/dH'$) extracted from the force curve. The margin of error less than 5% demonstrates the reliability and accuracy of the aforementioned calibration method. This method is further proved to be feasible for eliminating the topographic crosstalk in a-$Ga_2O_3$/ZnO nanowire array, 2D $NbOI_2$ flake and lotus leaf. This work not only reveals the underlying mechanism of topographic crosstalk in CAFM, but also introduces the comprehensive calibration method to optimize the accuracy of CAFM characterization.

## EXPERIMENTAL SECTION

**Fabrication of mica with uneven topography.** The pit on the mica substrate with the thickness of 200 µm were prepared with a UV photolithography procedure (URE-2000/35). The photoresist was spin-coated on mica substrate and baked for two minutes. Photolithography was then performed under UV light for eleven seconds. Finally, the non-light exposed part of the photoresist was etched using developer for obtaining the uneven-topography mica.

**Methods for controlling probe height in topography mapping.** The height of probe during the topography mapping was adjusted by controlling the force applied to the sample. When positive force was set, the probe remained in contact with the sample (white part in Figure S1). Conversely, when negative force was set, the probe elevated until it reached its maximum height (black part in Figure S1). The calibration of AFM probe was achieved through the thermal noise method, which allowed for precise determination of both the inverse optical lever sensitivity (InvOLS) and spring constant of the probe. After calibration, the deflection of the probe determines the force applied to the sample (force = spring constant × InvOLS × (preset deflection − initial deflection)).[20]

The "Litho" module in Asylum Research software (version 15.09112) was utilized to achieve spatial profile mapping of applied force by modifying the "ScanMaster.ipf" program.[21] Simultaneously, the ORCA (CAFM) module's current amplifier was used to measure the corresponding current flow through the probe.

**Methods for controlling probe height in force curve.** By setting dwell parameters in contact mode force curves, when the probe touched the sample and reached the predefined threshold force (trigger point), it moved along preset height, and continuously collected signals instead of lifting up immediately.

**Methods for calculating the first derivative of height.** To obtain the first derivative of the topographic map, we calculated the scanning time of each line. The scanning time of Figure 1h, 6b, 7a, 7d and 7g were 0.399 s, 0.102 s, 0.102 s, 0.113 s and 0.666 s, respectively. NumPy function was used to calculate the first derivative of height with respect to time in Python.

**Characterization and Measurements.** CAFM measurements were performed using an atomic force microscope (Asylum Research MFP-3D origin+). The Ti/Ir coated conductive tips, model ASYELEC-01-R2, were utilized in ORCA mode (CAFM). The crystalline structure and topography characteristic were characterized by X-ray diffraction (XRD; Bruker D8 Advance) and scanning electron microscopy (SEM; Zeiss Geminisem500), respectively. Optical microscopy images were characterized by dry transfer technique (Metatest E1-T).

**Fabrication of a-$Ga_2O_3$/ZnO nanowire array.** The ZnO seeds layer were deposited onto the ITO bottom electrode at a substrate temperature of 400 °C via radio frequency (RF) magnetron sputtering. Then the ZnO nanowire array was grown by using hydrothermal method, and the growth solution was prepared based on the previous work.[7] Finally, the amorphous $Ga_2O_3$ (a-$Ga_2O_3$) layer were deposited on ZnO nanowire array with 300 °C via RF magnetron sputtering.

**Fabrication of 2D $NbOI_2$ flake.** The $NbOI_2$ single crystals were grown by using chemical vapor transport (CVT) method. The starting materials Nb, $Nb_2O_5$ and $I_2$ powders with a ratio of 3:1:6 were sealed in an evacuated quartz tube. The tube was ramped to 700 °C within 2 h, and maintained at this temperature for three days. The thin $NbOI_2$ flakes were mechanically exfoliated and transferred onto a fluorine-doped tin oxide (FTO) substrate through a dry transfer technique (Metatest E1-T).

## ASSOCIATED CONTENT

**Supporting Information.**

The preset deflection image used to control the height of the probe during scanning; the discrepancy of Figure 1b,e; the $dI_{total}/dH'$ and corresponding errors of five force curves (PDF); the topography and crystalline structure characterizations and corresponding schematic diagrams of CAFM tests; the cross-sectional lines extracted from Figure 7b,c,e,f,h,i.

## ACKNOWLEDGEMENTS

This work was supported by the National Natural Science Foundation of China (No. 52372107) and the Natural Science Foundation of Henan Province in China (No. 212300410004 and No. 222300420125).

# GRAPHICAL ABSTRACT (TOC)

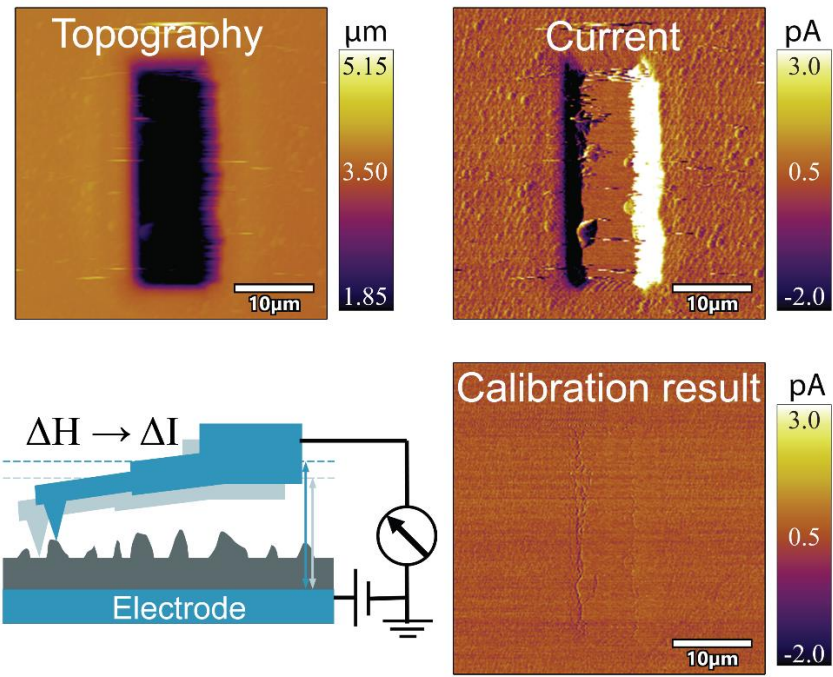

# Supporting information

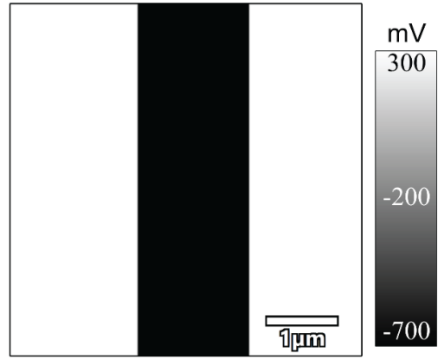

**Figure S1.** The preset deflection controlling the height of the probe during CAFM scanning.

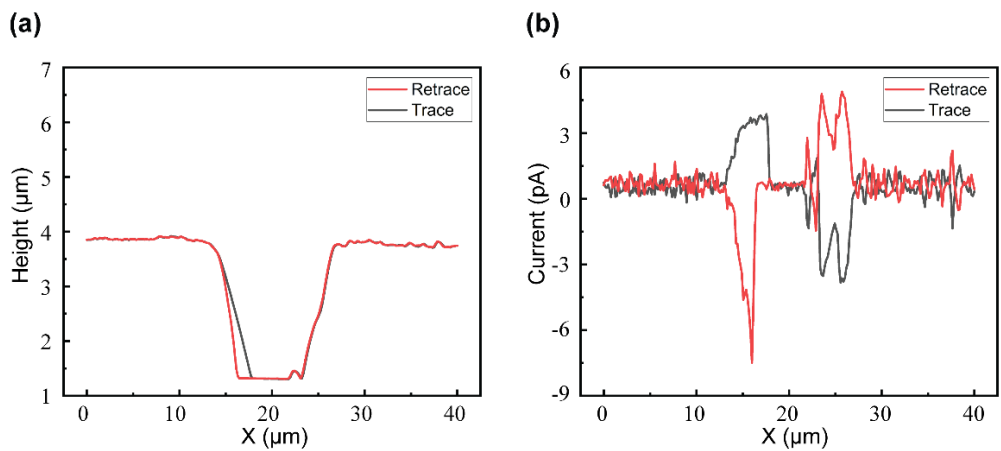

**Figure S2.** Cross-sectional lines extracted from (a) Figure 1b,e and (b) Figure 1d,g at the same position.

As illustrated in Figure S2a, the height trace does not coincide with the retrace, which results from various factors, such as scanning rate, I gain and probe model. As a result, there is the difference in corresponding current trace and retrace (Figure S2b).

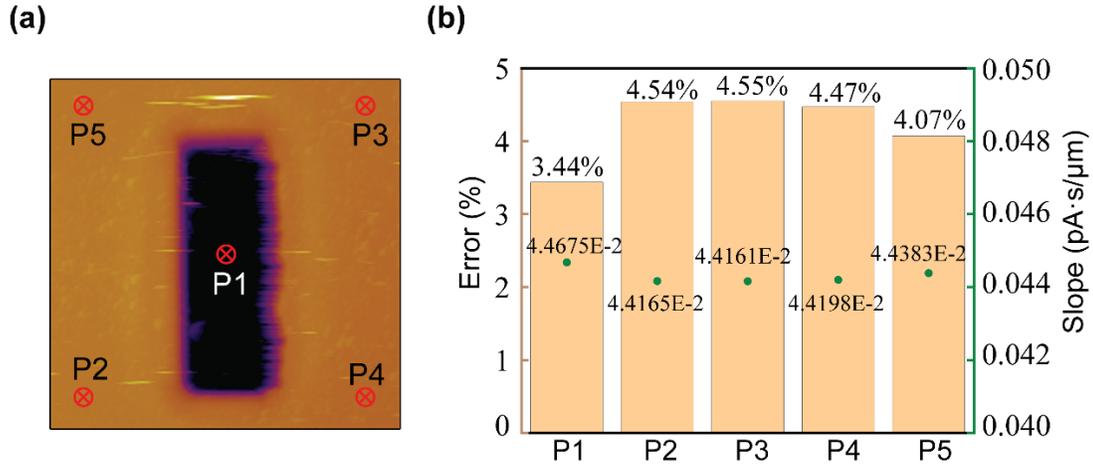

**Figure S3.** (a) Locations of the five force curves. (b) The d$I_{total}$/d$H'$ and corresponding errors of five force curves.

To assess the accuracy of updating d$I_{total}$/d$H'$ from the force curve other than topographic mapping, five force curves were collected at distinct locations in Figure S3. The d$I_{total}$/d$H'$ derived from Figure 5c was regarded as the standard value and labeled as $K_0$. The d$I_{total}$/d$H'$ extracted from force curves was labeled as $K$. The $K$ values extracted from the force curve and topographic map are positive and negative in AR software, respectively. The error can be expressed as:

$$Error = \left| \frac{|K| - |K_0|}{K_0} \right|$$

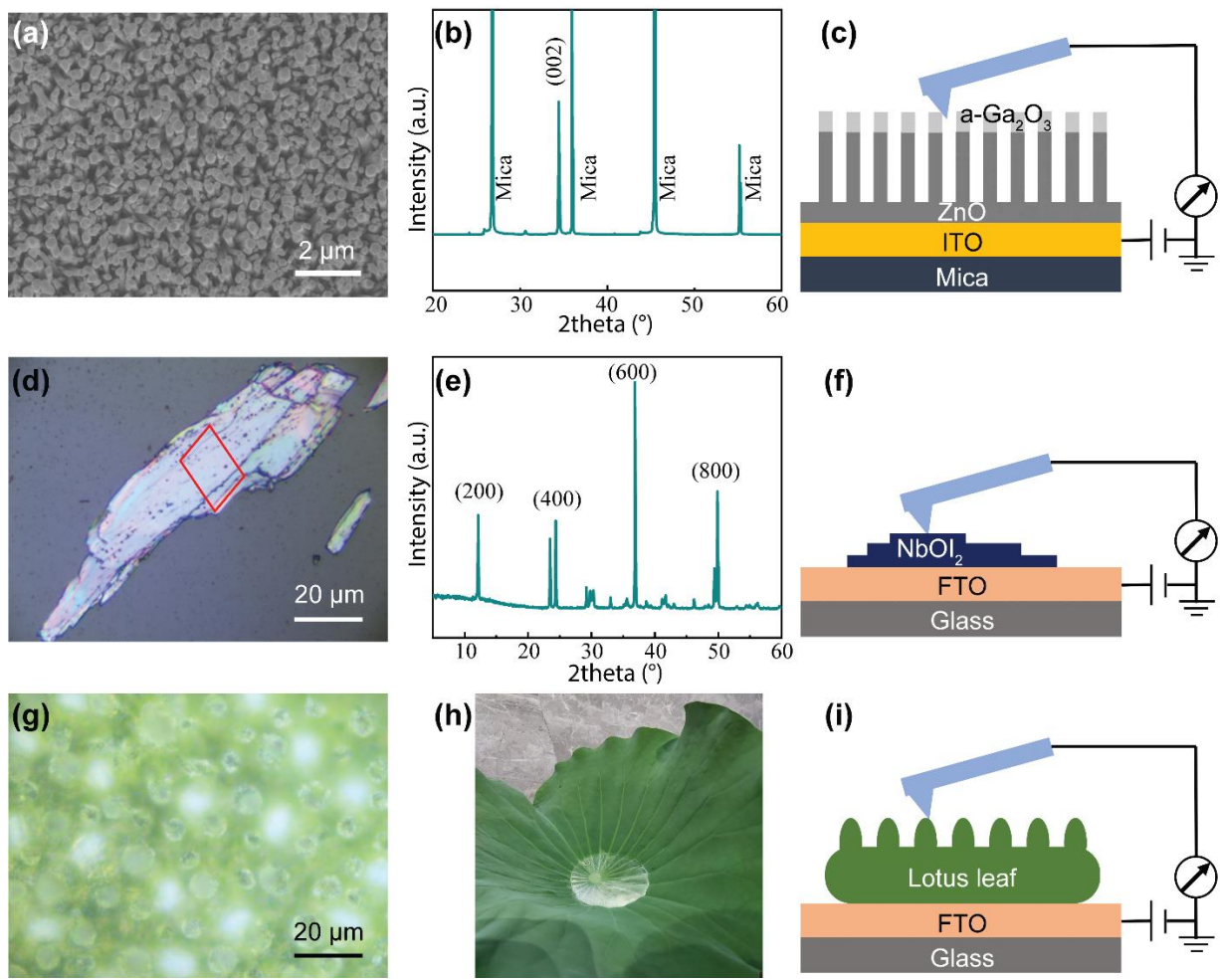

**Figure S4.** (a) SEM image of a-Ga$_2$O$_3$/ZnO nanowire array. (b) XRD pattern of ZnO nanowire array. (c) Schematic diagram of CAFM test on a-Ga$_2$O$_3$/ZnO nanowire array. (d) Optical microscopy image of 2D NbOI$_2$ flake (red region corresponding to Figure 7d-f). (e) XRD pattern of the NbOI$_2$ powders. (g) Photograph of superhydrophobic lotus leaf. (h) Schematic diagram of CAFM test on 2D NbOI$_2$ flake. (g) Schematic diagram of CAFM test on lotus leaf. (i) Optical microscopy image of lotus leaf.

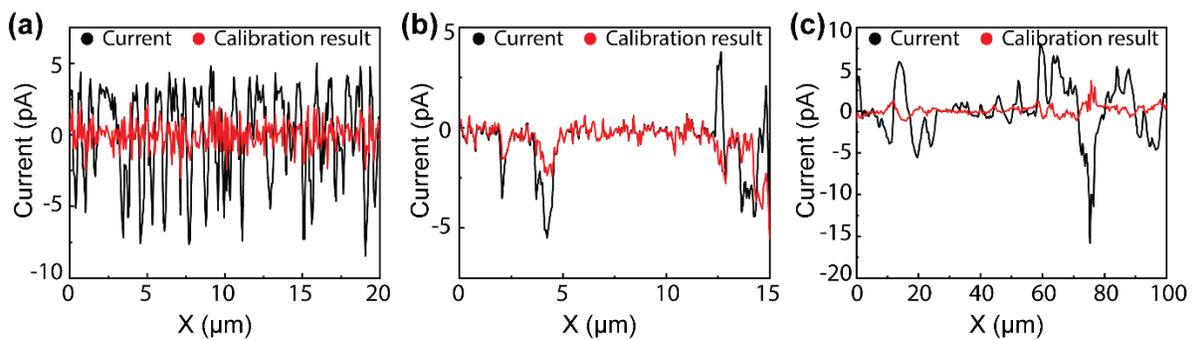

**Figure S5**. Cross-sectional lines extracted from (a) Figure 7b,c, (b) Figure 7e,f, (c) Figure 7h,i, respectively.